\documentclass[fleqn,usenatbib]{mnras}

\usepackage{xcolor}

\usepackage{siunitx}
\usepackage{amsmath}
\usepackage{subfigure}

\usepackage{placeins}

\usepackage{caption}

\usepackage{multirow}
\usepackage{graphicx}
\usepackage{txfonts}
\usepackage{stfloats}

\usepackage{mathtools} 
\usepackage{hyperref} 

\usepackage{natbib}
\bibpunct{(}{)}{;}{a}{}{,} 

\usepackage{url}

\newcommand{\deq}{\coloneqq} 
\newcommand{\bn}{\hat{\mathbf{n}}} 
\newcommand{\br}{\mathbf{r}} 

\usepackage[T1]{fontenc}
\usepackage{ae,aecompl}

\title[Tomographic constraints on the high-energy cosmic neutrino emission rate]{\textit{Tomographic constraints on the high-energy cosmic neutrino emission rate}}

\author[A. Gálvez Ureña  et al.]{Alberto Gálvez Ureña$^{1 ,2}$%
\thanks{Contact e-mail: \href{mailto:urena@fzu.cz}{urena@fzu.cz}}%
Federico Urban,$^{1}$
David Alonso$^{3}$
\\
$^{1}$CEICO, Institute of Physics of the Czech Academy of Sciences, Na Slovance 1999/2, 182 21 Prague, Czech Republic\\
$^{2}$Institute of Theoretical Physics, Faculty of Mathematics and Physics, Charles University,V Hole\v sovi\v{c}kách 2, 180 00 Prague 8, Czech Republic\\
$^{3}$Department of Physics, University of Oxford, Denys Wilkinson Building, Keble Road, Oxford OX1 3RH, United Kingdom}

\date{Accepted XXX. Received YYY; in original form ZZZ}

\pubyear{{\the\year{}}}

\begin{document}
\label{firstpage}
\pagerange{\pageref{firstpage}--\pageref{lastpage}}
\maketitle

\begin{abstract}
 
Despite growing efforts to find the sources of high energy neutrinos measured by IceCube, the bulk of the neutrinos remain with unknown origins. In this work, we aim to constrain the emissivity of cosmic high-energy neutrinos from extragalactic sources through their correlation with the large-scale structure. We use cross-correlations between the IceCube 10-year dataset and tomographic maps of the galaxy overdensity to place constraints on the bias-weighted high-energy neutrino emissivity out to redshift $z\sim3$. We test two different models to describe the evolution of neutrino emissivity with redshift, a power law model $\propto (1+z)^a$, and a model tracking the star formation history, assuming a simple power law model for the energy injection spectrum. We also consider a non-parametric reconstruction of the astrophysical neutrino emissivity as a function of redshift. We do not find any significant correlation, with our strongest results corresponding to a $1.9 \sigma$ deviation with respect to a model with zero signal. We use our measurements to place upper bounds on the bias-weighted astrophysical high-energy neutrino emission rate as a function of redshift for different source models. This analysis provides a new probe to test extragalactic neutrino source models. With future neutrino and galaxy datasets we expect the constraining and detection power of this type of analysis analysis to increase.
\end{abstract}

\begin{keywords}
High Energy Neutrinos, Harmonic Cross-Correlation, IceCube, Large Scale Structure
\end{keywords}

%

\section{Introduction}
\label{sec:intro}

The identification of the sources responsible for the diffuse flux of high-energy astrophysical neutrinos, first detected by the IceCube Observatory \citep{IceCube:2013cdw}, is an open problem in contemporary astrophysics. Although specific sources or source classes have been associated with observed high-energy neutrinos \citep{IceCube:2018dnn,IceCube:2022der,IceCube:2023ame}, these identifications collectively account for only a minor component of the total measured flux. The overwhelming majority of these high-energy events exhibit a distribution of arrival directions consistent with large-scale isotropy \citep{IceCube:2020acn}, thereby leaving the characteristics of the dominant source populations largely undetermined.
Searching for anisotropies in the celestial distribution of neutrino arrival directions is among the most promising methods to identify the underlying neutrino source population. Astrophysical neutrinos are expected to be produced in extragalactic sources whose distribution traces that of the large-scale structure (LSS) of matter in the Universe. In this case, since the large-scale structure is anisotropic, the neutrino flux should carry an imprint of this anisotropy. In particular, this means that there should be a non-zero cross-correlation between the arrival directions of the neutrinos and the distribution of matter in the Universe, for instance as traced by galaxies  \citep{Fang:2020rvq, Ouellette:2024ggl} or the unresolved gamma-ray sky \citep{Negro:2023kwv}. Statistical anisotropies are in particular useful when direct source association is challenging to perform due to the low astrophysical neutrino purity among large amount of neutrinos detected.

The angular cross-correlation between cosmic messengers and galaxy catalogues has emerged as a powerful tool to determine the properties of the sources of gamma rays \citep{Cuoco:2015rfa}, ultra-high-energy cosmic rays \citep{Urban:2020szk} and gravitational waves \citep{Raccanelli:2016cud}. The measurement of the cross-power spectrum between a neutrino sky map and a galaxy survey permits a statistical detection of a correlation between the two fields, which in turn can be used to constrain the characteristics of the neutrino source population, even when the emission from any individual source is below the threshold for direct detection.
In this work we develop a cross-correlation analysis in order to measure the product of the neutrino source bias and the comoving neutrino emissivity (or comoving neutrino energy emissivity). In order to do so we utilise ten years of IceCube muon-track data \citep{IceCube:2021xar} and four broad galaxy and quasar catalogues: the 2MASS Photometric Redshift catalogue (2MPZ; \citealp{Bilicki:2013sza}), the WISExSuperCOSMOS catalogue (WIxSC; \citealp{Bilicki:2016irk}), a sample from the DESI Legacy Imaging Surveys (DECaLS; \citealp{Hang:2020gwn}), and the Gaia-unWISE quasar catalogue (Quaia; \citealp{Storey-Fisher:2023gca}). These catalogues collectively cover the broad redshift range $z \lesssim 3$. The resulting measurements are interpreted in terms of two distinct phenomenological models for the unknown redshift evolution of the source population -- for which we can derive constraints on their model parameters -- or in a model-independent way by taking advantage of the tomographic coverage of our galaxy catalogues.

Our analysis builds upon the cross-correlation studies of \citet{Fang:2020rvq,Ouellette:2024ggl}, but differs in that we focus on the physical parameters that define the absolute neutrino flux, in particular the comoving neutrino emissivity and the comoving neutrino energy emissivity, together with the linear bias between neutrinos and the large-scale structure. Our method allows us to employ the cross-correlation to obtain a measure of these physical parameters, which has not previously been reported in the literature.

The paper is organised as follows: in \autoref{sec:theory} we define the neutrino flux via the comoving neutrino emissivity and the comoving neutrino energy emissivity, and describe the angular, harmonic cross-correlation between neutrinos and galaxies; in \autoref{sec:data} we introduce the neutrino and galaxy data and detail how we build the neutrino and galaxy maps from these data; in \autoref{sec:results} we present our analysis and our results. Finally, in \autoref{sec:conclusion} we summarise our results and give an outlook for future work.

\section{Theory}
\label{sec:theory}

\subsection{Neutrino intensity and emissivity}
\label{ssec:intensity}
We start by defining the neutrino intensity $I_N(\bn)$ from a position $\bn$ in the sky as the number of neutrinos observed per unit time, energy, detector area, and solid angle (all measured in the observer's frame):
\begin{equation}
    I_N \deq \frac{dN_o}{d\varepsilon_odt_odA_od\Omega_o} \,.
\end{equation}
Considering sources emitting from a range of comoving distances $d\chi$ around $\chi$, we can relate the intensity $I_N(\bn)$ to the neutrino emissivity $j_N$ (the number of neutrinos emitted per unit energy, time, and volume in the emitter's frame, also referred to as the neutrino production rate density) as
\begin{equation}
    dI_N \deq \frac{j_N(\chi\bn,z,\varepsilon_e)}{4\pi(1+z)^3}d\chi
    \deq j_{N,c}(\chi\bn,z,\varepsilon_e)\frac{d\chi}{4\pi} \,,
\end{equation}
where $j_{N,c}$ is the comoving emissivity. Quantities labelled $o$ or $e$ are in the observer or emitter frames respectively. Assuming that the neutrinos only lose energy because of cosmological redshift, the line-of-sight integral version of this relation is is
\begin{equation}
    I_N(\bn,\varepsilon_o)=\int \frac{d\chi}{4\pi}j_{N,c}(\chi\bn,z,\varepsilon_o(1+z)) \,.
\end{equation}
In practice, the quantity we measure is the intensity integrated over a given energy band $\varepsilon_\mathrm{min}\leq\varepsilon\leq\varepsilon_\mathrm{max}$:
\begin{equation}
    {\cal I}_N(\bn)\deq \int_{\varepsilon_\mathrm{min}}^{\varepsilon_\mathrm{max}}d\varepsilon_o\,I_N(\bn,\varepsilon_o) \,.
\end{equation}

\subsection{Modelling the emissivity}
\label{ssec:emissivity}

In order to model the emissivity we use a simple model in which all sources have the same spectrum $s(\varepsilon_e)$, which we can write as
\begin{equation}
    \frac{dE}{dt_e\,d\varepsilon_e} \deq L_\nu\,s(\varepsilon_e) \,,
\end{equation}
where $L_\nu$ is the total energy emitted in the form of neutrinos per unit time, such that $\int d\varepsilon_e\,s(\varepsilon_e)=1$. Given a comoving luminosity function for the sources $dn_c/dL_\nu$, one can express the comoving neutrino emissivity as
\begin{equation}
    j_{N,c}(\br,z,\varepsilon_e) = \frac{s_\nu(\varepsilon_e)}{\varepsilon_e}\int dL_\nu\,L_\nu\,\frac{dn_c}{dL_\nu}(\br,z,L_\nu) \,,
\end{equation}
where $\br \deq \chi\bn$. Modelling the energy spectrum as a power law spectrum with spectral index $\beta$ (i.e.\ $s_\nu(\varepsilon_e)\propto\varepsilon\,_e^\beta$), we can express everything in terms of the observed energy as
\begin{equation}
    {\cal I}_N(\bn) = \int\frac{d\chi}{4\pi}\,\frac{\dot{n}_\nu(\chi\bn,z)}{(1+z)^{1-\beta}} \,,
\end{equation}
where we have introduced the comoving emission density rate of neutrinos integrated over the energy band
\begin{align}
    \dot{n}_\nu(\br,z) &\deq \int_{\varepsilon_\mathrm{min}}^{\varepsilon_\mathrm{max}} d\varepsilon_o\,j_{N,c}(\br,z,\varepsilon_o) \nonumber\\
    &= \int_{\varepsilon_\mathrm{min}}^{\varepsilon_\mathrm{max}} d\varepsilon_o\,\frac{s_\nu(\varepsilon_o)}{\varepsilon_o}\,\int dL_\nu\,L_\nu\,\frac{dn_c}{dL_\nu}(\br,z,L_\nu) \, .
    \label{eq:dotn}
\end{align}
Given the three-dimensional matter overdensity $\delta(\chi\bn,z)$ we can write $\dot{n}_\nu(\br,z) \deq \dot{\bar{n}}_\nu(z)\,[1+b_\nu\,\delta(\br,z)]$, where $\dot{\bar{n}}_\nu(z)$ is the average comoving neutrino number density rate, and $b_\nu$ is the luminosity-weighted bias of neutrino-emitting sources
\begin{equation}
  b_\nu\deq\frac{\int dL_\nu\,L_\nu(d\bar{n}_c/dL_\nu)\,b(L_\nu)}{\int dL_\nu\,L_\nu(d\bar{n}_c/dL_\nu)},
\end{equation}
with $b(L_\nu)$ the linear bias for sources with luminosity $L_\nu$.

\subsection{Energy weights}
\label{ssec:en_weights}

In the previous section we have assumed that what we measure is the number of neutrinos in a given energy band (and from a given direction). However, since we can also measure the energy of each neutrino, we can also measure the total energy in neutrinos per unit time, solid angle, detector area and neutrino energy:
\begin{equation}
    I_E \deq \frac{dE_o}{d\varepsilon_o\,dt_o\,dA_o\,d\Omega_o} = \varepsilon_o\,I_N \,.
\end{equation}
Introducing the comoving energy emissivity (also referred to as the neutrino luminosity density) $j_{E,c}(\varepsilon_e) \deq \varepsilon_e\,j_{N,c}(\varepsilon_e) = s_\nu(\varepsilon_e)\int dL_\nu\,L_\nu\,\frac{dn_c}{dL_\nu}(\br,z,L_\nu)$, we find
\begin{equation}
    I_E(\bn)=\int\frac{d\chi}{4\pi}j_{E,c}(\chi\bn,z,\varepsilon_o(1+z)) \,.
\end{equation}
Similarly to the previous subsection, we integrate in energy and assume a power law with spectral index $\beta$ to finally obtain
\begin{equation}
    {\cal I}_E(\bn)= \int_{\varepsilon_\mathrm{min}}^{\varepsilon_\mathrm{max}} d\varepsilon_o\,I_E(\bn,\varepsilon_o)=\int\frac{d\chi}{4\pi}\frac{\dot{\rho}^b_\nu(\chi\bn,z)}{(1+z)^{-\beta}} \,,
\end{equation}
where we defined the comoving neutrino energy density rate as
\begin{align}
    \dot{\rho}_\nu(\br,z) &\deq \int_{\varepsilon_\mathrm{min}}^{\varepsilon_\mathrm{max}}d\varepsilon_o\,j_{E,c}(\br,z,\varepsilon_o) \nonumber\\
    &= \int_{\varepsilon_\mathrm{min}}^{\varepsilon_\mathrm{max}} d\varepsilon_o\,s_\nu(\varepsilon_o)\,\int dL_\nu\,L_\nu\,\frac{dn_c}{dL_\nu}(\br,z,L_\nu) \,.
    \label{eq:dotrho}
\end{align}
Just as before, this quantity can be related to the three-dimensional matter overdensities as $\dot{\rho}_\nu(\br,z) \deq \dot{\bar{\rho}}_\nu(z)\,[1+b_\nu\,\delta(\br,z)]$.

\subsection{Harmonic-space cross-correlations}
\label{ssec:cross}

Given a matter tracer $U$ defined on a two-sphere, we can define its two-dimensional anisotropies $\Delta U$ as a projection of the three-dimensional matter overdensities $\delta(\chi\bn,z)$:
\begin{equation}
    \Delta U(\bn) \deq \int d\chi\,q_U(\chi)\,[b_{U}\bar{u}](z)\,\delta(\chi\bn,z) \,,
    \label{eq:nu_od}
\end{equation}
where, for the neutrino fluxes above, $q_U(\chi)$ is the radial kernel $q_U(\chi) \deq [4\pi(1+z)^{1-\beta}]^{-1}$ and $\bar{u}$ is either $\dot{\bar{n}}_\nu$ or $\dot{\bar{\rho}}_\nu$.

The second matter tracer that we want to correlate with the neutrinos is a catalogue of galaxies, for which, in analogy with \autoref{eq:nu_od}, we can write as
\begin{equation}
    \Delta_g(\bn) \deq \int d\chi\,q_g(\chi)\,b_g\,\delta(\chi\bn,z) \,,
\end{equation}
where the radial kernel in this case is given by
\begin{equation}
    q_g(\chi) \deq H(z)\,\frac{dp}{dz} \,,
\end{equation}
$b_g$ is the galaxy bias and $dp/dz$ is the redshift distribution of galaxies in the catalogue.

In the Limber approximation which is valid for broad kernels such as those we employ here \citep{LoVerde:2008re}, the cross-correlation between $\Delta U$ and $\Delta_g$ is
\begin{equation}
    C_\ell^{UG} \deq \int\frac{d\chi}{\chi^2}q_g(\chi)q_U(\chi)b_g\,[b_\nu\bar{u}](z)\,P(k_\ell,z) \,,
    \label{eq:xc_ug}
\end{equation}
with $P(k,z)$ being the matter power spectrum and the wavenumber $k_\ell$ is related to the harmonic multipole $\ell$ as $k_\ell\deq(\ell+1/2)/\chi$.

Assuming all bias parameters to be constant for simplicity, we can pull them out of the integrals in the cross-correlations to finally obtain
\begin{align}
    C_\ell^{{\cal I}_{N} \, G} &= b_g\int\frac{d\chi}{\chi^2}q_g(\chi)\frac{b_\nu\,\dot{\bar{n}}_\nu(z)}{4\pi(1+z)^{1-\beta}}\,P(k_\ell,z) \,,
    \label{eq:xc_ing} \\
    C_\ell^{{\cal I}_{E} \, G} &= b_g\int\frac{d\chi}{\chi^2}q_g(\chi)\frac{b_\nu\,\dot{\bar{\rho}}_\nu(z)}{4\pi(1+z)^{-\beta}}\,P(k_\ell,z) \,,
    \label{eq:xc_ieg}
\end{align}
for the number-based and energy-based neutrino maps, respectively. In the same way the galaxy autocorrelation is
\begin{equation}
    C_\ell^{G \, G}=b_g^2\int\frac{d\chi}{\chi^2}\,[q_g(\chi)]^2\,P(k_\ell,z) \,.
    \label{eq:ac}
\end{equation}

\subsection{Modelling the neutrino emission rate}\label{ssec:nu_distro}
As discussed in \autoref{ssec:emissivity}, the neutrino flux fluctuations depend on the luminosity function of the neutrino sources, as well as their evolution in time, their clustering properties, and the neutrino energy spectrum. However, the cross-correlation measurements are only sensitive to the final neutrino emission rates $\dot{\bar{n}}_\nu$ or $\dot{\bar{\rho}}_\nu$ (Eqs. \ref{eq:dotn} and \ref{eq:dotrho}), resulting from integrating the luminosity function, multiplied by the effective bias of the neutrino sources $b_\nu$. Since the nature of the astrophysical neutrino sources is not known, in this work we will use simple parametrisations to characterise the redshift dependence of the bias-weighted emission rates. In particular we will consider three models:
\begin{itemize}
  \item A {\bf power law model} of the form
  \begin{equation}
    b_\nu\dot{\bar{n}}_\nu(z) = Nb_\nu\,(1+z)^a \,; \quad b_\nu\dot{\bar{\rho}}_\nu(z) = N_\rho b_\nu\,(1+z)^a.\label{eq:nu_power}
  \end{equation}
  \item A {\bf peak model} tracking the star formation history. In this case:\begin{equation}\label{eq:nu_peak}
    b_\nu\dot{\bar{n}}_\nu(z) = Nb_\nu\,f_{\rm SFR}(z),\hspace{6pt}
    b_\nu\dot{\bar{\rho}}_\nu(z) = N_\rho b_\nu\,f_{\rm SFR}(z) \,,
  \end{equation}
  where $f_{\rm SFR}(z)$ is a redshift-dependent function that describes the evolution of the star-formation rate density $\rho_{\rm SFR}$ as proposed by \cite{Madau_2014}:
  \begin{equation}
    f_{\rm SFR}(z)\equiv\frac{(1+z)^{2.7}}{1+((1+z)/2.9)^{5.6}}.
  \end{equation}
  \item A {\bf tomographic} model, in which the bias-weighted emission rates are characterised by their values at the mean redshifts of the four galaxy samples studied here, which we constrain from the separate analysis of each cross-correlation.
\end{itemize}
Here $N$ and $N_\rho$ are normalisation constants with units of ${\rm Mpc}^{-3}\,{\rm yr}^{-1}$ and ${\rm erg}\,{\rm Mpc}^{-3}\,{\rm yr}^{-1}$, characterising the late-time mean neutrino emissivity. Due to the degeneracy with the effective neutrino bias $b_\nu$, we will treat the combinations $Nb_\nu$ and $N_\rho b_\nu$ as free parameters. Because we do not assume a specific type of neutrino source, we do not have an expectation for $b_\nu$; however all but the rarest of astrophysical sources have biases in the range $1\lesssim b\lesssim 5$, and therefore $b_\nu$ should be a quantity of order ${\cal O}(1)$. Therefore, our constraints on $Nb_\nu$ or $N_\rho b_\nu$ will still allow us to place bounds on the order of magnitude of the global astrophysical neutrino emissivity. In the power-law model, the emissivity decays with redshift at a rate characterised by an additional parameter $a$ (assuming $a$ to be negative). In turn, the peak model follows the star formation history, peaking at $z\sim2$.

\section{Data}
\label{sec:data}

\subsection{High-energy neutrinos}
\label{ssec:nu_map}
\begin{figure}
    \centering
    \begin{subfigure}
        \centering
        \includegraphics[width=\hsize]{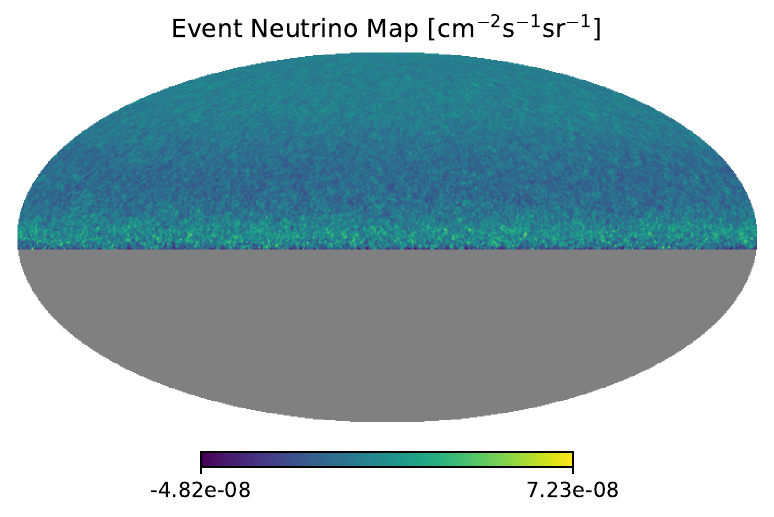}
    \end{subfigure}

    \begin{subfigure}
        \centering
        \includegraphics[width=\hsize]{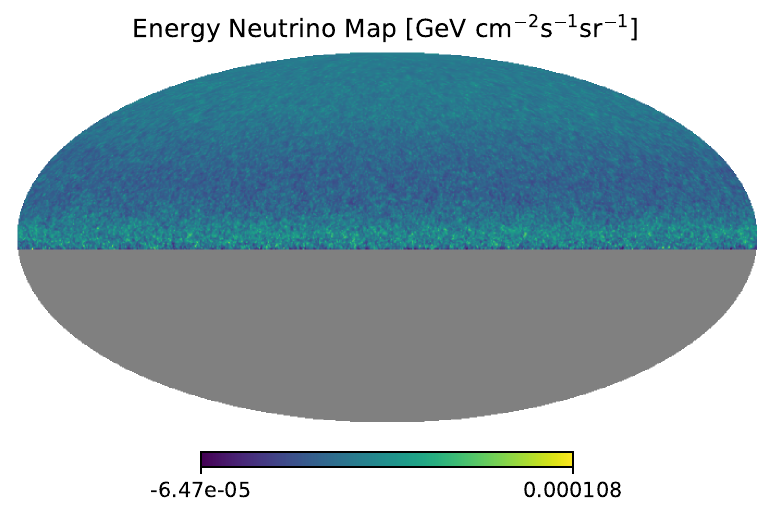}
    \end{subfigure}
    \caption{Neutrino intensity maps. The top panel displays the number-density-based map, the bottom panel displays the energy-density-based map (in GeV units).}
    \label{fig:nu_maps}
\end{figure}

In our analysis we used the 10-year IceCube point source data set \citep{IceCube:2021xar}. This data set is composed of track-like events in which a muon is detected. The muon produces a long track in the detector, from which we can obtain the arrival direction of the muon with an angular resolution of about $\ang{1}$.

The IceCube data set is divided in 10 seasons, each of them with its own uptime, effective area and smearing matrix files. This last one is a five-dimensional matrix that relates the energy and angular error of the reconstructed muon to the true energy, declination and angular separation to the muon of the original neutrino. For this work we will neglect the effects of the smearing matrix, effectively assuming that the parameters of the reconstructed muon represent well the original neutrino. We expect this to be a good approximation when it comes to energy, since we are considering wide energy bins. At the energies considered in this work, the effect on the angular directions is expected to be similar to the angular resolution of the muons. We then expect this angular correction to have a small impact on the analysis, since large angular scales contribute the most due to the beaming (see effective beam below). The effective area is interpolated to give a continuous function $A_{\rm eff}(\delta,\varepsilon_o)$. To be precise, the effective area data is binned in energy and declination. For each bin \(b\), take the value of the effective area to correspond to an energy $E_b = \sqrt{E_{b-}\,E_{b+}}$ and a declination $\delta_b = 0.5\left( \delta_{b-}+\delta_{b+} \right)$, where \(+\) and \(-\) indicate the upper and lower limits of the bin respectively. Then, using a grid of points $E_b$ and $\sin{\delta_b}$, we interpolate linearly to obtain $A_{\rm eff}(\delta,\varepsilon_o)$. We obtain the neutrino maps for each season $i$ separately. Each event contributes $1/A_{{\rm eff}}$ to the number intensity map ${\mathcal{I}}^{i}_{N}(\bn)$ and $\varepsilon_{o}/A_{{\rm eff}}$ to the energy intensity map ${\mathcal{I}}^{i}_{E}(\bn)$. We then assign a weight to each event according to a two-dimensional Gaussian filter with width corresponding to the angular uncertainty for each event. Considering all events as circular Gaussians is an approximation, since more detailed reconstructions often have non-gaussian distributions\cite{HESE}. Note that we exclude all events with effective areas below $10\,{\rm cm}^2$ to avoid numerical artefacts---this is only relevant for events with high declinations and energies over $10^5$~GeV.

The total number intensity and energy intensity maps are obtained by merging the seasonal maps according to
\begin{equation}
    {\mathcal{I}}(\bn)= \frac{\sum_{i=1}^{10}\,\bar{A}_{\rm eff}^i(\bn)\,{\mathcal{I}}^{i}(\bn)}{\Omega_{\mathrm{pix}}\sum_{i=1}^{10}\, T_i\,\bar{A}_{\rm eff}^i(\bn)} \,.
    \label{eq:data2}
\end{equation}

In this expression we define $\Omega_{pix}$ as the angular size of each pixel, $T_i$ as the uptime of season $i$ and $\bar{A}_{\rm eff}^i(\bn)$ as the effective area averaged over all energies for season $i$. The latter one we obtain by dividing the energy range into small energy bins and assuming that in each bin the neutrinos follow a power law with spectral index $\alpha$. Let $(\varepsilon_n,\varepsilon_{n+1})$ be the edges of the $n$-th interval, with $\varepsilon_1 \deq \varepsilon_\mathrm{min}$ and $\varepsilon_{N+1} \deq \varepsilon_\mathrm{max}$ the minimum and maximum values respectively. Then the average effective area for each pixel is calculated as
\begin{equation}
    \bar{A}_{{\rm eff},p}^i = \frac{1}{{\varepsilon_{\rm max}^{\alpha+1}-\varepsilon_{\rm min}^{\alpha+1}}}\sum_{n=1}^NA_{\rm eff}(\delta_p,\frac{\varepsilon_{n+1} - \varepsilon_n}{2})\,[\varepsilon_{n+1}^{\alpha+1}-\varepsilon_n^{\alpha+1}] \,.
    \label{eq:data3}
\end{equation}
For this work we choose $\alpha = - 3.7$ corresponding to the tilt of atmospheric neutrinos \citep{IceCube:2014slq}. Note that \autoref{eq:data3} is only used to weight seasons against each other in a semi-optimal way. One could skip this step and the analysis would still be valid, but less optimal in terms of the noise properties of the coadded map.

Not all detected muons in a given energy range are coming from astrophysical neutrinos. Indeed, the largest muon contribution is atmospheric muons, that is, muons that are generated as secondary particles in the cosmic ray interactions in the atmosphere of the Earth. Given the location of IceCube, taking events with a declination larger than $-\ang{5}$ allows us to screen many of these atmospheric muons. The second largest muon contribution is from atmospheric (rather than astrophysical) neutrinos. Atmospheric neutrinos originate from cosmic ray interactions in the atmosphere; these neutrinos produce muons that are then detected. We can not avoid this background, but we know that the fraction of astrophysical (atmospheric) neutrinos increases (decreases) with energy. Hence, for our analysis we focus on the highest energies in the range $\varepsilon_\mathrm{min} = 10^3$~GeV to $\varepsilon_\mathrm{max} = 10^6$~GeV, for a total of 403,529 neutrinos. Note that the atmospheric neutrinos do not correlate with the LSS, so they will not bias our measurements, but they will contribute to the statistical uncertainties of the cross-correlation.

Once the final map is constructed, we compute the effect of finite angular resolution following the fiducial estimate of the effective beam described in \cite{Ouellette:2024ggl}. In order to do so we compute the effective beam function $B_\ell^{\mathrm{eff}} \deq C_\ell^{\bar{\nu}\bar{\nu}}/C_\ell^{\nu\nu}$, where $C_\ell^{\bar{\nu}\bar{\nu}}$ is the auto-correlation for the map with the individual Gaussian beams and $C_\ell^{\nu\nu}$ is the same for the neutrino map without any smoothing. We then will multiply \autoref{eq:xc_ing} and \autoref{eq:xc_ieg} by $B_\ell^{\mathrm{eff}}$ before comparing with the data. The finite resolution of IceCube limits the range of scales over which significant information can be extracted to $\ell\lesssim100$ (see Fig.~4 in \cite{Ouellette:2024ggl}).

In this work we will build neutrino maps with all neutrinos within the energy range above as well as maps of neutrinos binned in energy in order to constrain the astrophysical neutrino emission rate as a function of observer-frame energy, and to quantify the stability of our results against the presence of atmospheric neutrinos, which are more prominent at low energies.

Since we expect the angular distribution of atmospheric events to be dominated mostly by very large-scale features, we subtract the harmonic-space monopole (namely, the average over the map) and dipole from all our neutrino maps. This reduces the impact of the atmospheric neutrino contamination on the statistical uncertainties of the power spectra at higher \(\ell\)s, arising from mode-coupling due to the sky mask. 

Lastly, since we are interested in the physical values of the neutrino number density rate, \autoref{eq:dotn}, and energy density rate, \autoref{eq:dotrho}, unlike \cite{Ouellette:2024ggl} we will not normalise our neutrino maps by dividing by the monopole. For completeness, when integrated over the full energy range, the value of the monopole is $3.97 \cdot 10^{-8} \, \mathrm{cm}^{-2}\mathrm{s}^{-1}\mathrm{sr}^{-1}$ for the neutrino number flux map and $5.80 \cdot 10^{-5} \, \mathrm{GeV}\,\mathrm{cm}^{-2}\mathrm{s}^{-1}\mathrm{sr}^{-1}$ for the energy flux map, respectively. The final maps are shown in \autoref{fig:nu_maps}.

\subsection{Galaxy catalogues}
\label{ssec:galcat}
\begin{figure}
    \centering
    \includegraphics[width=\hsize]{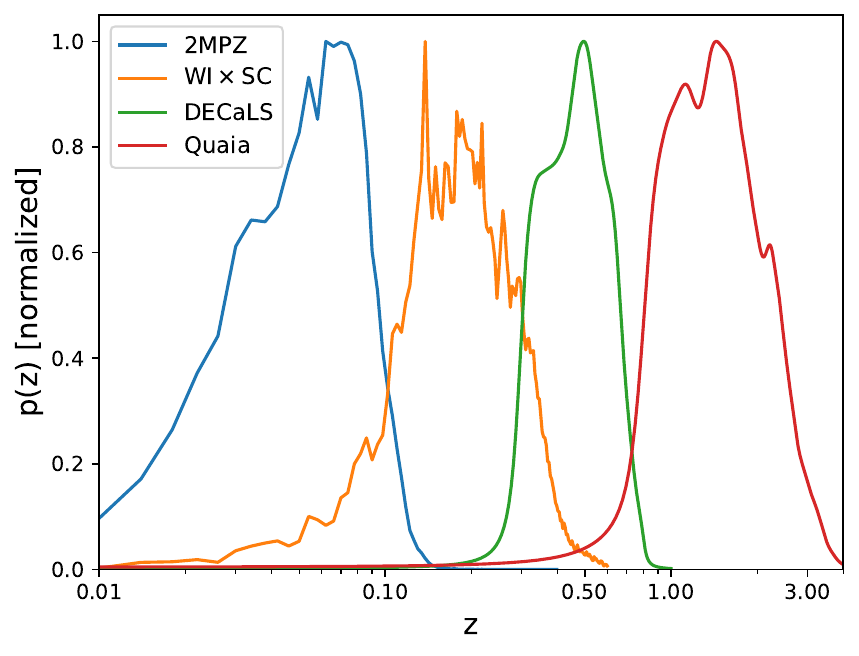}
    \caption{Redshift distributions of the four galaxy catalogues used here.}
    \label{fig:kernels}
\end{figure}

\begin{figure*}
    \centering
    \includegraphics[width=0.45\textwidth]{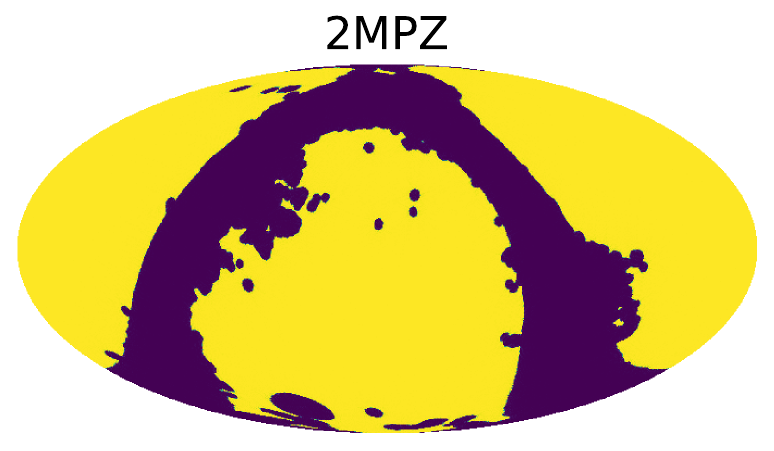}
    \includegraphics[width=0.45\textwidth]{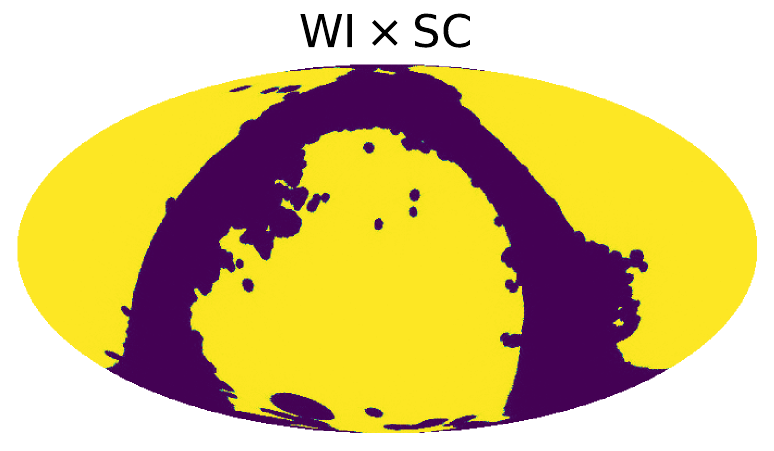}
    \includegraphics[width=0.45\textwidth]{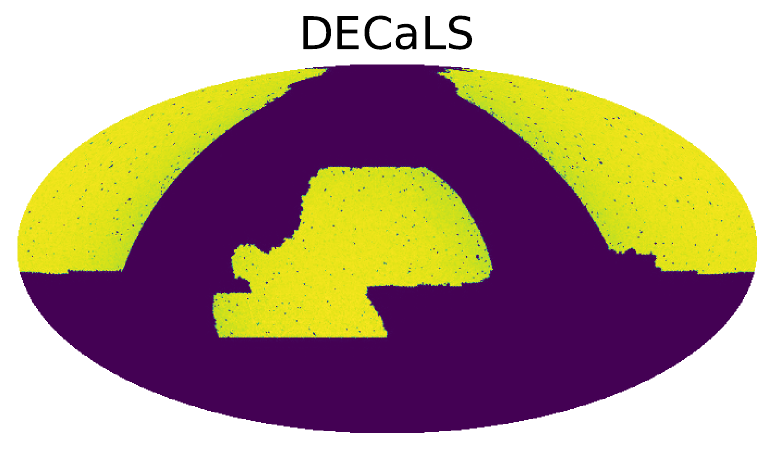}
    \includegraphics[width=0.45\textwidth]{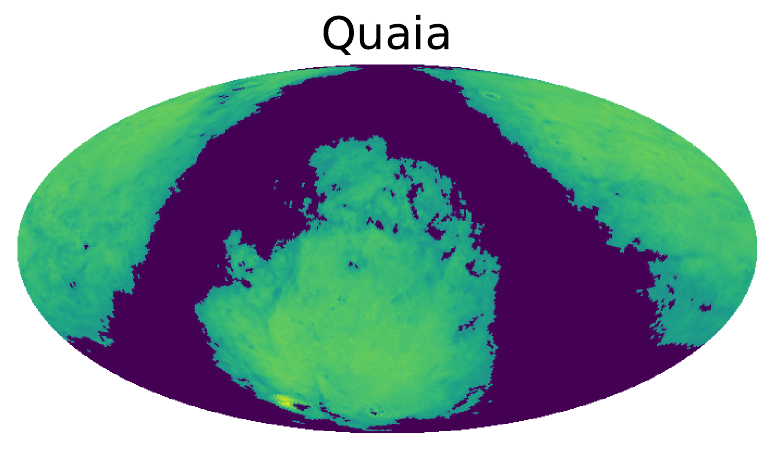}
    \caption{Sky coverage of the four different galaxy catalogues used.}
    \label{fig:gal_masks}
\end{figure*}

In order to cover a broad range of redshifts, and to capture the possible redshift evolution of the neutrino sources, we use four different galaxy catalogues. For each catalogue we restrict ourself to a range of complementary redshifts in order to minimise correlations between different bins, as explained in \autoref{sec:results} (see also \citet{Alonso:2024knf} for further details).

The first sample is the 2MASS Photometric Redshift catalogue 2MPZ \citep{Bilicki:2013sza}. The sample was selected by cross-matching the 2MASS \citep{Jarrett:2000me}, WISE \citep{Wright:2010qw} and SuperCOSMOS \citep{Peacock:2016wjw} all-sky surveys. A neural network approach is used to derive photometric redshifts with a precision of $\sigma_z = 0.015$, with the median redshift being of $z_m = 0.08$. For our sample, we select galaxies with photometric redshift $z < 0.1$, which gives us a sample with mean redshift $z_{av} = 0.064$ and 476,190 galaxies in total. For this catalogue we will also employ the mask described in \citet{Koukoufilippas:2019ilu} to remove the galactic plane and other regions contaminated by stars or dust.

The second sample is the WISE $\times$ SuperCOSMOS catalogue: WI$\times$SC \citep{Bilicki:2016irk}. This catalogue is obtained by cross-matching WISE and SuperCOSMOS. It has a median redshift of $z_m = 0.2$ and a redshift normalised scatter of $\sigma_z = 0.033$. We apply the same corrections as described in \citet{Koukoufilippas:2019ilu}. For our work we choose a sample with photometric redshift $0.1 < z < 0.4$ with mean redshift of $z_{av} = 0.23$, which consists of 16,325,449 galaxies.

The third catalogue is built from the DESI Legacy Imaging Surveys: DECaLS \citep{DESI:2018ymu}. This is a combination of the Dark Energy Camera Legacy Survey \citep{DES:2015wtr}, the Beijing-Arizona Sky Survey \citep{Zou_2019} and the Mayall $z$-band Legacy Survey \citep{Mayall}. Specifically, we use the sample selected by \citet{Hang:2020gwn} and apply their corrections for sky contamination. Our sample has a photometric redshift range of $0.3 < z < 0.8$, with an average of $z_{av} = 0.50$.

The last catalogue is the Gaia-unWISE quasar catalogue: Quaia \citep{Storey-Fisher:2023gca}. This catalogue is built from combining the Gaia quasar sample \citep{Gaia} and the infrared data from unWISE \citep{Meisner:2019lbf}. The sky contamination is corrected with a selection function as described in \citet{Storey-Fisher:2023gca}. Using this selection function we can build the sky mask as described in \citet{Alonso:2023guh}. Our selected sample has galaxies with $0.8 < z < 5$, with an $z_{av} = 1.72$, for a total of 1,092,207 quasars.

In \autoref{fig:kernels} we show the redshift distributions for the four galaxy catalogues after corrections to the photometric redshifts, normalised to unit peak. The redshift distributions for the two low-redshift samples were estimated via direct calibration, using cross-matched spectroscopic samples weighted in colour space (details can be found in \cite{Paopiamsap_2024}). In turn the redshift distributions of the two high-redshift samples were estimated by stacking the photo-$z$ probability distributions for all galaxies in the sample (see \cite{Hang:2020gwn} and \cite{Alonso:2023guh} for details about the DECaLS and Quia samples, respectively). The combination of all 4 catalogues allow us to cover the full redshift range out to $z\sim3$, with limited overlap between samples. In  \autoref{fig:gal_masks} we show the sky coverage of each survey in equatorial coordinates, only declinations above $-\ang{5}$ are useful for us due to the constrains on the neutrino map.

All the neutrino and galaxy maps are built with HEALPix \citep{Gorski:2004by} as implemented by \texttt{healpy} \citep{Zonca:2019vzt}.\footnote{\url{http://healpix.sf.net}} We use $\text{N}_\text{side} = 256$, corresponding to a pixel size of $\delta\theta\simeq0.22^\circ$. This is sufficient, given the limited angular resolution of the neutrino observations ($\delta\theta_\nu\sim0.7^\circ$. The dipole component of the neutrino maps was subtracted using the \texttt{remove\_dipole} method implemented in \texttt{healpy}.

\section{Methodology and results}
\label{sec:results}

To place constraints on the neutrino emissivity from cross-correlations with the galaxy catalogues described above, we use a tomographic approach consisting of three stages.
\begin{itemize}
    \item In the first step we compute the galaxy power spectrum and the cross-power spectrum between the galaxy overdensity and the neutrino intensity maps, as well as their covariance matrix. For this we employ the pseudo-$C_\ell$ estimator as implemented in \texttt{NaMaster} \citep{Alonso:2018jzx}, in order to properly take into account the partial sky coverage of all our data sets.
    \item We then estimate the theory power spectra from \autoref{eq:xc_ing}, \autoref{eq:xc_ieg} and \autoref{eq:ac} while varying the free parameters of each model (see \autoref{ssec:nu_distro}). The theory spectra are obtained with the Core Cosmology Library \citep{LSSTDarkEnergyScience:2018yem}.\footnote{\url{https://github.com/LSSTDESC/CCL}} We assume a flat $\Lambda$CDM cosmology with parameters $\Omega_c=0.25$, $\Omega_b = 0.05$, $\Omega_k = 0$, $\sigma_8 = 0.81$, $n_s = 0.96$, $h = 0.67$ and no massive neutrinos.
    \item In the last step we determine the free parameters by comparing the theory power spectra with the data through a likelihood analysis, which we describe below.
\end{itemize}

\subsection{Galaxy bias}\label{ssec:gal_bias}
To estimate the galaxy bias parameter $b_g$ for each of the samples used here, we use a Gaussian likelihood of the form
\begin{equation}
    \mathcal{L}(b_g) \deq -\frac{1}{2} \left(\mathbf{C}_D - \mathbf{C}_M(b_g)\right)^T \mathcal{M}^{-1} \left(\mathbf{C}_D - \mathbf{C}_M(b_g)\right) \,,
    \label{eq:gal_like}
\end{equation}
where the data vector $\mathbf{C}_D\deq C^{G\,G}_\ell$ contains our measurement of the galaxy power pectrum, and ${\bf C}_M(b_g)$ is the theoretical prediction for these measurements, dependent on $b_g$. $\mathcal{M}^{-1}$ is the inverse covariance matrix for $\mathbf{C}_D$. As described in section \ref{sec:theory}, we assume a linear bias model to describe the clustering of galaxies. This is only valid on relatively large scales, and therefore we only use multipoles in the range $\ell\leq k_{\rm max}\chi(\bar{z})$, where $\chi(\bar{z})$ is the comoving distance to the mean redshift of the sample, and $k_{\rm max}=0.15\,{\rm Mpc}^{-1}$. We assume a constant bias parameter for the 2MPZ, WI$\times$SC, and DECaLS samples. In the case of Quaia, given its broad redshift distribution, we account for the redshift evolution by assuming a redshift-dependent bias of the form  $b_\text{Quaia}(z) = b_g\,(0.278((1 + z)^2 - 6.565) + 2.393)$, where $b_g$ is a free parameter and the functional form is a numerical fit to the measured redshift evolution of quasar bias in \cite{Laurent_2017}.

Doing this, we determine the bias parameters of each sample to be:
\begin{align}
   &b_g^{\rm 2MPZ}=1.212 \pm 0.027,\hspace{12pt} b_g^{{\rm WI}\times{\rm SC}}=1.1608 \pm 0.0072\\
   &b_g^{\rm DECaLS}=1.4783 \pm 0.0059,\hspace{12pt}b_g^{\rm Quaia}=1.050 \pm 0.034.
\end{align}
A self-consistent analysis propagating the uncertainties in the measured galaxy bias when constraining the neutrino emission rate from the neutrino-galaxy cross-correlation would call for a joint likelihood combining both power spectra ($C_\ell^{GG}$ and $C_\ell^{{\cal I}_NG}$), simultaneously constraining $b_g$ and the neutrino emission properties. Nevertheless, given the high precision with which the linear bias of these samples is determined, we opt to simply treat them as fixed parameters in the analysis of the neutrino-galaxy cross-correlation, using the values above.

\subsection{Cross-correlation measurements}\label{ssec:xcorr}
\begin{figure}
    \centering
    \begin{subfigure}
        \centering
        \includegraphics[width=\hsize]{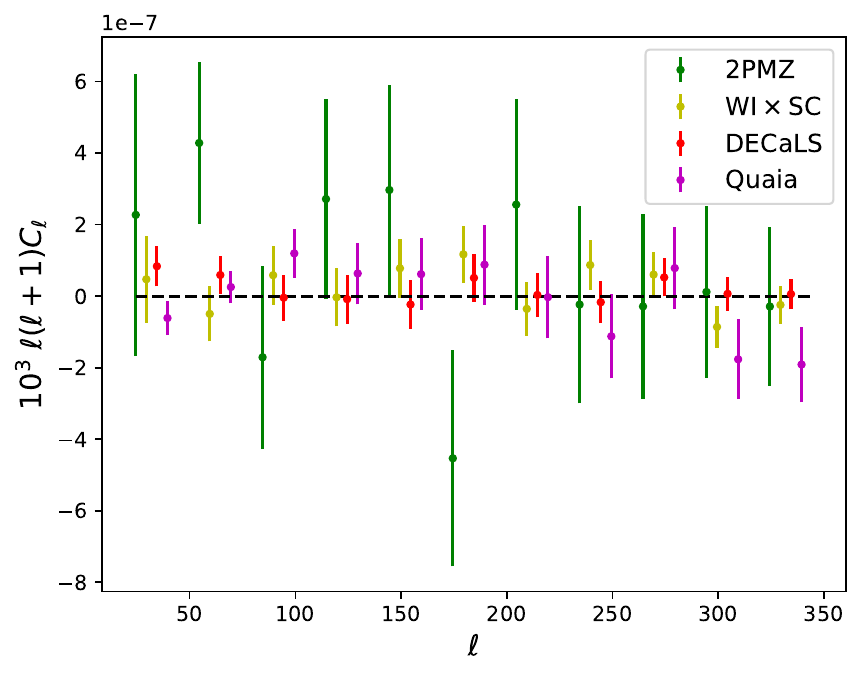}
    \end{subfigure}

    \begin{subfigure}
        \centering
        \includegraphics[width=\hsize]{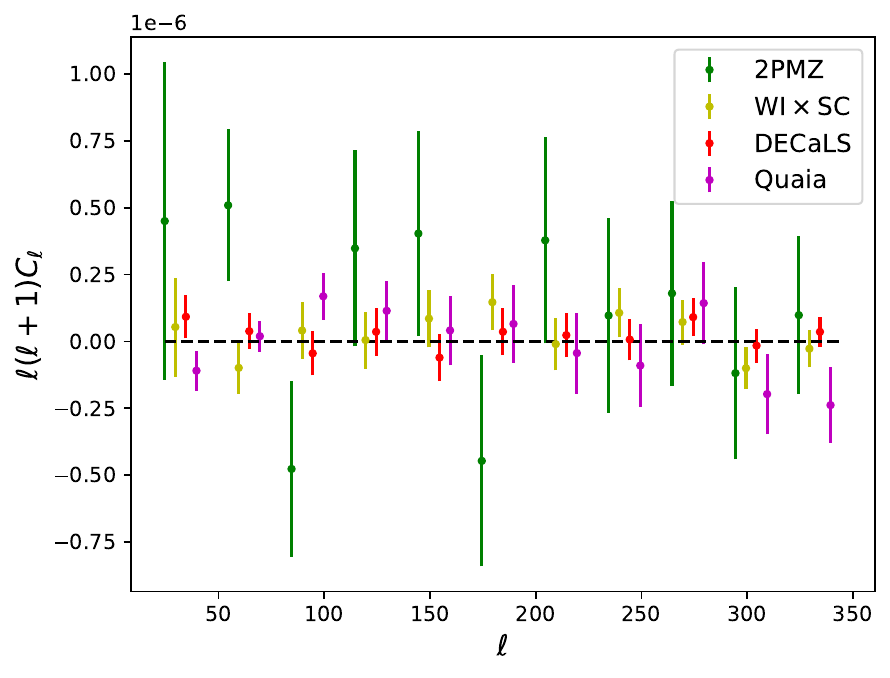}
    \end{subfigure}
    \caption{Top panel: number-based neutrino-galaxy cross-correlation for the galaxy catalogues 2MPZ (green), WI $\times$ SC (yellow), DECaLS (red) and Quaia (magenta). Bottom panel, energy-based neutrino-galaxy cross-correlations for the same catalogues. A dashed black line it shown at $C_\ell = 0$ for guidance. The points for different catalogues are all shifted in $\ell$ for illustration purposes, but they all correspond to the $\ell$ where the 2MPZ values are shown.}
    \label{fig:cross}
\end{figure}
Fig.~\ref{fig:cross} shows the cross-correlations between the neutrino number and energy flux maps (top and bottom panels, respectively), and the four different galaxy samples studied here. The contribution from atmospheric neutrinos is most prominent at low multipoles, significantly degrading the statistical uncertainties of the measurement. We therefore discard all data below multipole $\ell_\text{min} = 10$ in our analysis. Our measurements are binned into bandpowers with a constant width $\Delta\ell = 30$, and we use all measurements up to $\ell_\text{max} = 340$. Due to the limited resolution of the neutrino observations, no significant information can be gathered beyond this scale, as noted in \citep{Ouellette:2024ggl}. Furthermore, we have verified that the results reported here do not change significantly when varying the maximum multipole $\ell_\text{max}$.

To quantify the significance with which a cross-correlation signal is detected, we estimate its signal-to-noise ratio (SNR) as
\begin{equation}
    \text{SNR} = \text{sign}(S) \, \sqrt{|S|} \quad, \quad S = {\bf C}_D^T\mathcal{M}^{-1} {\bf C}_D-N_\text{dof},
\end{equation}
where ${\bf C}_D$ is a vector containing a given cross-correlation measurement (between a galaxy sample and either the number or energy flux maps), and $\mathcal{M}^{-1}$ is the inverse covariance matrix of this measurement. The number of degrees of freedom in this case is the number of data points (i.e. the size of ${\bf C}_D$), and accounts for the expected value of the null chi-squared $\chi^2_0\equiv{\bf C}_D^T\mathcal{M}^{-1} {\bf C}_D$ for purely noise-like data compatible with zero. The SNRs of the different samples studied here are listed in Table \ref{tab:snr_results}.
  \begin{table}
    \centering
    \begin{tabular}{|l| c | c|}
      \hline
      Galaxy catalogue & ${\rm SNR}_N$ & ${\rm SNR}_\rho$ \\ 
      \hline
      2MPZ &   $-1.26$ & $-0.51$ \\
      WI$\times$SC &  $-1.34$ & $-1.80$\\
      DECaLS &  $-2.38$ & $-2.46$\\
      Quaia & $\phantom{-}1.67$ & $\phantom{-}1.51$\\
      \hline
    \end{tabular}
    \vspace{2pt}
    \caption{Signal to Noise ratios obtained from the cross-correlation of the neutrino map with each galaxy catalogue. Here ${\rm SNR}_N$ and ${\rm SNR}_\rho$ are the SNRs for the number- and energy-weighted neutrino flux maps}
    \label{tab:snr_results}
\end{table}

As the table shows, we do not find evidence of a clear correlation between the IceCube neutrino maps and the LSS tracers studied here. This can be visually verified in Fig.~\ref{fig:cross}, where all cross-correlation measurements are compatible with zero correlation within the statistical uncertainties. This is in qualitative agreement with the results of \cite{Ouellette:2024ggl}, who did not find strong evidence of correlation between the neutrino map and the galaxy overdensity. Note, however, that the authors found tentative detections of a non-zero cross-correlation with some of their galaxy samples (at the level of $\sim2\sigma$ at most). It is worth noting that the methods used here differ significantly from those of \cite{Ouellette:2024ggl}, particularly their normalisation of the neutrino counts by the local mean density of neutrino events. This procedure may enhance the sensitivity of the resulting map to cross-correlations with LSS tracers, at the cost of losing information about the unnormalised amplitude of the neutrino anisotropies that do correlate with the LSS. Avoiding this is what allows us to translate our measurements into constraints on physical parameters describing the extragalactic neutrino emission rate.

\subsection{Constraining the neutrino emission rate}\label{ssec:constraints}
\begin{table}
    \centering
    \begin{tabular}{| l | c | c | c |}
    \hline
    Model & $Nb_\nu$ & $N_\rho b_\nu$ & $a$ \\
    \hline
    Power law 
    & $5.8 \pm 3.6$ 
    & $7.9 \pm 6.3$ 
    & $-2.7 \pm 1.4$ \\
    Peak
    & $4.0 \pm 2.1$ 
    & $4.7 \pm 3.6$ 
    & \\
    \hline
    \end{tabular}
    \vspace{2pt}
    \caption{$68\%$ confidence level constraints on the free parameters of the power law and peak models used to describe the neutrino emission rates (see \autoref{ssec:nu_distro}). The constraints on the amplitude parameters $Nb_\nu$ and $N_\rho b_\nu$ are given in units of $10^{45}\,\mathrm{Mpc}^{-3}\,\mathrm{yr}^{-1}$ and $10^{45}\,\mathrm{erg}\,\mathrm{Mpc}^{-3}\,\mathrm{yr}^{-1}$, respectively.} 
    \label{tab:results}
\end{table}

\begin{figure}
    \centering
    \includegraphics[width=0.49\textwidth]{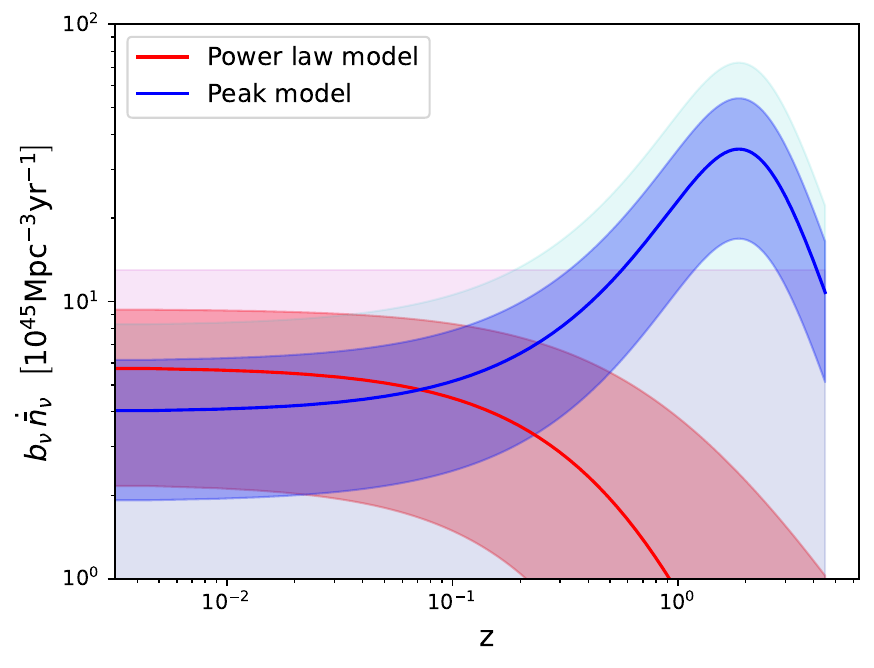}
    \caption{Allowed regions for the bias-weighted number density rate of neutrinos times in the two parametric models considered here (red and blue for the power law and peak models). We show the 1$\sigma$ and 2$\sigma$ constraints.}
    \label{fig:cont_constrains}
\end{figure}
To estimate the neutrino emissivity parameters from the measured cross-correlations we make use of a likelihood-based Bayesian inference approach. As before, we assume a Gaussian likelihood, now including data from multiple galaxy samples:
\begin{equation}
    \mathcal{L}(\vec{\theta}) \deq -\frac{1}{2}\sum_{G} \left(\mathbf{C}_{D,G} - \mathbf{C}_{M,G}(\vec{\theta})\right) (\mathcal{M}_G)^{-1} \left(\mathbf{C}_{D,G} - \mathbf{C}_{M,G}(\vec{\theta})\right).
    \label{eq:nu_like}
\end{equation}
As before, ${\bf C}_{D,G}$ is the measured galaxy-neutrino cross-correlation for the $G$-th galaxy sample, and ${\bf C}_{M,G}(\vec{\theta})$ is the theoretical prediction for this measurement, dependent on the model parameters $\vec{\theta}$. ${\cal M}_G$ is the covariance matrix of the measurements. We will consider cross-correlations with both the number-based and energy-based neutrino flux maps. The free parameters of the theory model are $\vec{\theta}\in\{a,Nb_\nu/N_\rho b_\nu\}$ or $\vec{\theta}\in\{Nb_\nu/N_\rho b_\nu\}$ for the power-law and peak neutrino sources redshift distribution models, respectively. For the analysis we assume $\beta = 2.5$ following \cite{IceCube:2024fxo}, which we use in \autoref{eq:xc_ing} and \autoref{eq:xc_ieg}. We sample this likelihood using the nested sampling Monte Carlo algorithm \texttt{MLFriends} \citep{Ultranest1,Ultranest2} as implemented in the \texttt{UltraNest} package \citep{Ultranest3}.\footnote{\url{https://johannesbuchner.github.io/UltraNest/}} We assume flat uninformative priors on the amplitude parameters: $Nb_\nu$ and $N_\rho b_\nu / \mathrm{erg} \in \left[-10^4,10^4\right]\cdot 10^{45}\,\mathrm{Mpc}^{-3}\mathrm{yr}^{-1}$, and we consider only negative values for the index of the power law model, with $a \in \left[ -5,0 \right]$.

We note that the form of the likelihood in Eq. \ref{eq:nu_like} assumes that there is negligible correlation between the cross-correlation measurements in the different redshift bins. This is not entirely correct, since the different galaxy samples have a non-zero overlap in their redshift distributions. This overlap is small, however (see Fig.~\ref{fig:kernels}), and thus we do not expect this assumption to affect our results significantly.
  
The $68\%$ constraints we obtain on the free parameters of the two parametric models considered here are summarised in \autoref{tab:results}. For the number-based intensity maps, the corresponding amplitude parameters are positive but compatible with zero at the $1.6\sigma$ and $1.9\sigma$ level for he power-law model and the peak model, respectively. This is comparable with the significance with which the overall amplitude of the cross-correlation was measured in \cite{Ouellette:2024ggl}. The evidence for a positive correlation is smaller for the energy-based intensity maps, with amplitude parameters consistent with zero at the $\sim1.3\sigma$ level. We note that the power law index $a$ is unconstrained within its prior for both the number-based and energy-based intensity maps. The bounds on the neutrino emissivity as a function of cosmic time resulting from these constraints, at 1 and 2$\sigma$, for the power law and peak models, are shown in \autoref{fig:cont_constrains}. We report only the constraints on the number-based emission rate, since the results for $b_\nu\dot{\rho}_\nu$ are effectively equivalent.
  \begin{figure}
    \centering
    \includegraphics[width=0.49\textwidth]{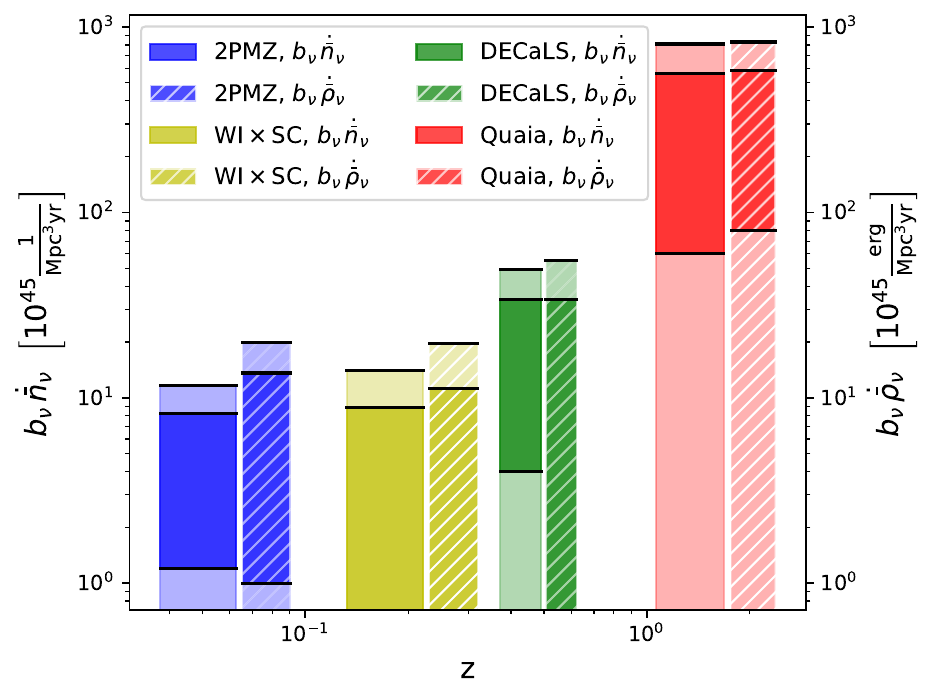}
        \caption{1$\sigma$ and 2$\sigma$ allowed regions for the number and energy density rates of neutrinos times $b_\nu$ when considering the galaxy catalogues individually under the assumption of constant neutrino density rate over each single bin.}
        \label{fig:cte_constraints}
\end{figure}

\begin{table}
    \centering
    \begin{tabular}{|l| c | c | c|}
    \hline
    & $z_\mathrm{av} \pm \sigma_z$  & $b_\nu\,\dot{\bar{n}}_\nu$ & $b_\nu\,\dot{\bar{\rho}}_\nu$ \\ 
    \hline
    2MPZ & $0.064 \pm 0.026$ &  $4.7 \pm 3.5$ & $7.3 \pm 6.3 $ \\
    WI$\times$SC & $0.23 \pm 0.09$ & $3.7 \pm 5.2$ & $2.8 \pm 8.4$\\
    DECaLS & $0.50 \pm 0.13$ & $19 \pm 15$ & $13 \pm 21$\\
    Quaia & $1.72 \pm 0.66$ & $310 \pm 250$ & $330 \pm 250$\\
    \hline
    \end{tabular}
    \vspace{2pt}
    \caption{$68\%$ confidence level tomographic constraints on both $b_\nu\,\dot{\bar{n}}_\nu$ (in units of $10^{45}\,\mathrm{Mpc}^{-3}\,\mathrm{yr}^{-1}$) and $b_\nu\,\dot{\bar{\rho}}_\nu$ (in units of $10^{45}\,\mathrm{erg}\,\mathrm{Mpc}^{-3}\,\mathrm{yr}^{-1}$), effectively assuming a constant value within each redshift bin. In the second column we show the average redshift of each bin together with the standard deviation of the associated redshift distribution for each galaxy catalogue.}\label{tab:cte_results}
\end{table}

In addition to constraining these two parametric models of the astrophysical neutrino emission rate, we make model-independent measurements using the tomographic approach described in \autoref{ssec:nu_distro}, assuming that the emission rate varies only slowly within each redshift bin considered here. Furthermore, these constraints are obtained by analysing the cross-correlation with galaxy sample independently from the rest. The results are shown in \autoref{fig:cte_constraints}, with the numerical constraints on $b_\nu\dot{\bar{n}}_\nu$ and $b_\nu\dot{\bar{\rho}}_\nu$ listed in \autoref{tab:cte_results}. As in the case of the parametric models considered above, although the measured values of the bias-weighted emission rate are consistently positive, both for $\dot{\bar{n}}_\nu$ and $\dot{\bar{\rho}}_\nu$, they are compatible with zero within $2\sigma$ in all cases.

To our knowledge, these are the first tomographic constraints on the emissivity of the diffuse astrophysical background of high-energy neutrinos.

  \subsection{Consistency tests}\label{ssec:consistency}
We performed a number of tests to evaluate the consistency and robustness of the results presented in the previous section.
    
First, to enable a more direct comparison with \cite{Ouellette:2024ggl}, we obtained constraints assuming a redshift dependence of the neutrino emissivity modelled after the expected distribution of galaxies from the Vera Rubin telescope \cite{LSSTScience:2009jmu}, peaking at $z\approx1$:
\begin{equation}
    b_\nu\dot{\bar{n}}_\nu(z) = Nb_\nu\,f_R(z);\quad
    b_\nu\dot{\bar{\rho}}_\nu(z) = N_\rho b_\nu\,f_R(z),
\end{equation}
with $f_R(z)\equiv(1+z)^2\exp(-z^2)$. In this case we find $Nb_\nu = (4.3 \pm 2.4) \cdot 10^{45}\ \mathrm{Mpc}^{-3}\,\mathrm{yr}^{-1}$ and $N_\rho b_\nu = (5.1 \pm 4.1)\cdot 10^{45}\ \mathrm{erg}\,\mathrm{Mpc}^{-3}\,\mathrm{yr}^{-1}$. The amplitude in the case of the number-based intensity map is compatible with zero within $2\sigma$, in qualitative agreement with \cite{Ouellette:2024ggl}, who find $b_\nu f_{\rm astro}=0.26\pm0.13$ for this model (where $f_{\rm astro}$ is the fraction of astrophysical neutrinos in the IceCube data). As with our other models, the energy-based intensity map recovers an amplitude that is more compatible with zero (within $1.2\sigma$).

As a robustness test against the impact of atmospheric neutrinos on our conclusions, we repeated the analysis binning the neutrino data set into three energy bins: $10^3\,\mathrm{GeV}\leq \varepsilon_o <10^4\,\mathrm{GeV}$, $10^4\,\mathrm{GeV}\leq \varepsilon_o <10^5\,\mathrm{GeV}$ and $\varepsilon_o \geq10^5\,\mathrm{GeV}$. The resulting constraints for these three energy bins are listed in \autoref{tab:results_app}. Despite the fact that atmospheric neutrinos are expected to be more dominant at lower energies, the significance of any potential detection of a cross-correlation with the LSS does not improve at higher energies (rather the opposite, in fact). It is worth noting that, at these higher energies the loss of statistics due to the significantly smaller dataset, may significantly degrade the possibility of detecting any existing correlations. For context, in the IceCube dataset analysed here, we find 393,804 events with $10^3\,\mathrm{GeV}\leq \varepsilon_o <10^4\,\mathrm{GeV}$, 9,521 events with $10^4\,\mathrm{GeV}\leq \varepsilon_o <10^5\,\mathrm{GeV}$ and 204 events with $\varepsilon_o \geq10^5\,\mathrm{GeV}$ above the declination cut of $-\ang{5}$.

Additionally, we also test two alternative declination cuts, keeping only data above $\delta \geq +\ang{0}$ and $\delta \geq +\ang{5}$. The results are again listed in \autoref{tab:results_app}. The results obtained using our default declination cut do not change qualitatively: no significant evidence of a cross-correlation with the galaxy overdensity is observed and, in fact, the best-fit amplitude parameters decrease marginally.

Finally, following the recent work by the IceCube collaboration \citep{abbasi2025constraintscorrelationicecubeneutrinos}, we generate an atmospheric neutrino template map by averaging the neutrino map in $50$ declination bins of equal $\sin(\delta)$. We then subtract this template from the neutrino map and repeat the analysis. We find no significant difference in this case (in fact, we find a very modest improvement). By correlating the atmospheric template map with galaxies we see that the atmospheric template only contributes to the lowest multipoles $\ell\lesssim20$; owing to this result, we repeat the analysis shifting the multipole range by a decade: $\ell \in \left[20 - 350\right)$. In this case we observe a slight decrease in the amplitude of the best-fit parameters. This is expected, since most of the constraining power comes from small $\ell$ due to the shape of the effective beam. We collect these results in \autoref{tab:results_app}.

Overall, these tests show that our results are robust with respect to the different methods used to minimise the impact of the unknown atmospheric contamination. This allows us to confidently interpret our results as an at best marginal detection of a correlation between the IceCube high-energy neutrinos and the low-redshift LSS. Nevertheless, by examining the cross-correlation with the unnormalised neutrino intensity maps, we are able to place upper bounds on physical parameters that quantify the expected background neutrino emissivity from extragalactic sources.

\begin{table*}
  \centering
  \begin{tabular}{| l | l | c | c | c |}
  \hline
  Model & Case
  & $Nb_\nu$ $\left[10^{45}\,\mathrm{Mpc}^{-3}\,\mathrm{yr}^{-1}\right]$
  & $N_\rho b_\nu$ $\left[10^{45}\,\mathrm{erg}\,\mathrm{Mpc}^{-3}\,\mathrm{yr}^{-1}\right]$
  & $a$ \\
  \hline
  & $10^3 \, \mathrm{GeV} \leq \varepsilon_o < 10^4 \, \mathrm{GeV}$
  & $\phantom{-}5.6 \pm 3.3$ 
  & $\phantom{-}8.0 \pm 5.8$ 
  & $-2.6 \pm 1.4$ \\
  & $10^4 \, \mathrm{GeV} \leq \varepsilon_o < 10^5 \, \mathrm{GeV}$
  & $-0.018 \pm 0.021$ 
  & $-0.48 \pm 0.45$ 
  & $-2.7 \pm 1.4$ \\
  & $\varepsilon_o \geq 10^5 \, \mathrm{GeV}$
  & $\phantom{-}0.00009 \pm 0.00031$ 
  & $\phantom{-}0.064 \pm 0.064$ 
  & $-2.5 \pm 1.4$ \\
  Power law & $\delta \geq +\ang{0}$
  & $\phantom{-}4.1 \pm 3.4$ 
  & $\phantom{-}5.2 \pm 6.3$ 
  & $ -2.7 \pm 1.4 $ \\
  & $\delta \geq +\ang{5}$
  & $\phantom{-}1.4 \pm 3.4$ 
  & $\phantom{-}1.8 \pm 6.4$ 
  & $-2.7 \pm 1.4$ \\
  & Substract atm.\ template 
  & $5.6 \pm 3.1$
  & $8.3 \pm 5.6$
  & $-2.8 \pm 1.4$ \\
  & $\ell \in \left[20 - 350\right)$ 
  & $5.1 \pm 3.2$
  & $7.7 \pm 5.8$
  & $-2.8 \pm 1.4$ \\
  \hline
  & $10^3 \, \mathrm{GeV} \leq \varepsilon_o < 10^4 \, \mathrm{GeV}$
  & $\phantom{-}4.0 \pm 2.1$ 
  & $\phantom{-}4.9 \pm 3.5$ 
  & \\
  & $10^4 \, \mathrm{GeV} \leq \varepsilon_o < 10^5 \, \mathrm{GeV}$
  & $-0.011 \pm 0.013$ 
  & $-0.28 \pm 0.25$ 
  & \\
  & $\varepsilon_o \geq 10^5 \, \mathrm{GeV}$
  & $\phantom{-}0.00012 \pm 0.00020$
  & $\phantom{-}0.054 \pm 0.038$ 
  & \\
  Peak  & $\delta \geq +\ang{0}$
  & $\phantom{-}2.7 \pm 2.1$ 
  & $\phantom{-}3.6 \pm 3.7$ 
  & \\
  & $\delta \geq +\ang{5}$
  & $\phantom{-}1.1 \pm 2.1$ 
  & $\phantom{-}2.4 \pm 3.9$
  &\\
  & Substract atm.\ template 
  & $3.5 \pm 1.8$
  & $4.2 \pm 3.1$
  & \\
  & $\ell \in \left[20 - 350\right)$ 
  & $3.4 \pm 2.0$
  & $4.2 \pm 3.2$
  & \\
  \hline
  \end{tabular}
  \vspace{2pt}
  \caption{$68\%$ confidence interval for the parameters characterising the neutrino emission rate in the two source models explored here. The results listed correspond to the different consistency tests carried out to quantify the impact of atmospheric neutrinos on our measurements. These correspond to the analysis of neutrino maps in different energy ranges, and the use of different declination cuts.}\label{tab:results_app}
\end{table*}

\section{Discussion and conclusion}
\label{sec:conclusion}
In this work we have computed the angular, harmonic cross-correlation between the IceCube 10-year point source neutrino dataset and galaxies, in order to test whether neutrino sources positively correlate with the large-scale structure. Our analysis, leveraging a tomographic approach with four distinct galaxy catalogues spanning the redshift range $0 < z \lesssim 3$, yields constraints on the clustering and emissivity properties of high-energy neutrino sources. We do not find sufficiently compelling evidence for a positive correlation between the neutrino intensity maps and the large-scale structure, with all cross-correlations being compatible with zero within 1-2$\sigma$. We introduce two simple models to describe the neutrino source distribution while keeping the number of free parameters to a minimum, and constrain the amplitude of the neutrino emission rate to be $b_\nu \dot{\bar{n}}_\nu = 5.8\pm3.6\times10^{45}\,\mathrm{Mpc}^{-3}\,\mathrm{yr}^{-1}$ and $b_\nu \dot{\bar{n}}_\nu = 4.0\pm2.1\times10^{45}\,\mathrm{Mpc}^{-3}\,\mathrm{yr}^{-1}$ for the power-law and peak models, respectively. Both models produce similarly significant results for the comoving number density rate of neutrinos, not reaching the  $2 \sigma$ level (see \autoref{tab:results}). These results allow us to constrain the physical density rate as reflected in \autoref{fig:cont_constrains}. In the power law case we also attempt to estimate the power law index $a$, but the data is not sufficiently sensitive to constrain it beyond its prior.

We also take advantage of the range of redshifts covered by the galaxy samples used here to determine the redshift dependence of the bias-weighted neutrino emission rate in a tomographic and model-independent. The results are shown in \autoref{tab:cte_results}, and allow us to constrain the emission rate from $b_\nu \dot{\bar{n}}_\nu = (4.7\pm3.5)\times10^{45}\,\mathrm{Mpc}^{-3}\,\mathrm{yr}^{-1}$ for the low-redshift 2MPZ catalogue, at $z\simeq0.06$, to $b_\nu \dot{\bar{n}}_\nu = (310\pm250)\times10^{45}\,\mathrm{Mpc}^{-3}\,\mathrm{yr}^{-1}$ for the high-redshift Quaia sample, at $z\simeq1.7$. Although the mean emissivity grows with redshift, the statistical uncertainties grow in tandem, and we do not find evidence of a positive correlation in any redshift bin beyond $2\sigma$.

A minor shortcoming of this analysis is our treatment of the large-scale galaxy bias as a fixed, known quantity, given its relatively small uncertainties derived from the galaxy auto-correlation. As the sensitivity of neutrino observatories increases, and if a detection of the correlation between the neutrino signal and the large-scale structure is made, fully propagating all parameter uncertainties will become of vital importance. In this case, this could be easily achieved through a joint analysis of the galaxy auto-correlation and the galaxy-neutrino cross-spectra analysed here.

These findings are consistent with and complementary to previous searches for a neutrino-large-scale-structure correlation. Compared to the analyses of \citet{Fang:2020rvq} and \citet{Ouellette:2024ggl}, our method can determine the physical parameters describing the neutrino flux, namely the number emissivity $\dot{\bar{n}}_\nu$ and energy emissivity $\dot{\bar{\rho}}_\nu$, albeit only in combination with the effective bias $b_\nu$ of astrophysical neutrino sources. This is made possible by constructing neutrino maps that preserve the absolute flux scale, which can thus be measured directly in the event of a significant cross-correlation detection. Our results can be juxtaposed against the expectations from several classes of neutrino sources \citep{Murase_2016,Groth:2025aan}, which however suffer from large uncertainties in both the overall neutrino luminosity and the radial kernels of the sources.

Next-generation facilities such as the neutrino observatories IceCube-Gen2 and KM3NeT combined with the depth and sky coverage of upcoming galaxy surveys such as the Vera C. Rubin Observatory (LSST) and Euclid will significantly improve the sensitivity of cross-correlations between neutrinos and galaxies. This will increase the potential to make a detection of this correlation, improve our understanding of astrophysical neutrino sources and emission mechanisms, and potentially use neutrinos as a cosmological probe. Until then, further improvements will depend on a more robust modelling of the atmospheric flux, which would allow us to more reliable subtract this dominant source of contamination.
  
Measuring the neutrino emissivity as a function of redshift would provide a direct estimate of the total neutrino luminosity density of the Universe at a given epoch. This can be compared with theoretical predictions for various candidate source populations such as star-forming galaxies, specific classes of active galactic nuclei, or tidal disruption events, thereby providing a powerful tool to discriminate between candidates for the origin of high-energy neutrinos.

\section*{Acknowledgments}
The Authors thank Mauricio Bustamante and Stefano Camera for their contribution at the early stage of this project. AGU and FU acknowledge support from the European Structural and Investment Funds and the Czech Ministry of Education, Youth and Sports (project No.\ FORTE---CZ.02.01.01/00/22\_008/0004632). AGU also acknowledges the support of Charles University through GAUK, project No. 258623. DA acknowledges support from the Beecroft Trust.


\section*{Data Availability}

The data generated for this article will be shared on reasonable request with the corresponding authors. The underlying data is all publicly available, see references in text.

\label{lastpage}

\bibliographystyle{mnras}
\bibliography{main}  

\end{document}